\documentclass[letterpaper, 10 pt, conference]{ieeeconf}
\IEEEoverridecommandlockouts{}
\usepackage{amsmath}
\usepackage{amsfonts}
\usepackage{mathtools}
\usepackage{biblatex}
\usepackage{algorithm}

\usepackage{xcolor}

\usepackage{algorithmicx}
\usepackage{algpseudocode}
\newcommand{\R}{\mathbb{R}}
\newcommand{\PI}{\mathbf{\Pi}}
\bibliography{_lib/cpssecurity.bib, _lib/dynamicalsystems.bib}
\DeclareMathOperator*{\argmin}{\arg{} \,\min}
\title{\LARGE \bf
Efficient Predictive Monitoring of Linear Time-Invariant Systems Under Stealthy Attacks
}

\author{Mazen Azzam$^{1}$, Liliana Pasquale$^{2}$, Gregory Provan$^{3}$, and Bashar Nuseibeh$^{1}$
\thanks{$^{1}$Mazen Azzam ({\tt\small mazen.azzam@ul.ie}) and Bashar Nuseibeh ({\tt\small bashar.nuseibeh@ul.ie}) are with Lero, The Irish Software Research Centre, University of Limerick, Limerick, Ireland.}%
\thanks{$^{2}$Liliana Pasquale ({\tt\small liliana.pasquale@ucd.ie}) is with Lero, University College Dublin, Dublin, Ireland.}%
\thanks{$^{3}$Gregory Provan ({\tt\small g.provan@cs.ucc.ie}) is with Lero, University College Cork, Cork, Ireland.}%
}

\makeatletter
\newcommand{\removelatexerror}{\let\@latex@error\@gobble}
\makeatother

\begin{document}

\maketitle
\thispagestyle{empty}
\pagestyle{empty}


\begin{abstract}

Attacks on Industrial Control Systems (ICS) can lead to significant physical damage. While offline safety and security assessments can provide insight into vulnerable system components, they may not account for stealthy attacks designed to evade anomaly detectors during long operational transients. In this paper, we propose a predictive online monitoring approach to check the safety of the system under potential stealthy attacks. Specifically, we adapt previous results in reachability analysis for attack impact assessment to provide an efficient algorithm for online safety monitoring for Linear Time-Invariant (LTI) systems. The proposed approach relies on an offline computation of symbolic reachable sets in terms of the estimated physical state of the system. These sets are then instantiated online, and safety checks are performed by leveraging ideas from ellipsoidal calculus. We illustrate and evaluate our approach using the Tennessee-Eastman process. We also compare our approach with the baseline monitoring approaches proposed in previous work and assess its efficiency and scalability. Our evaluation results demonstrate that our approach can predict in a timely manner if a false data injection attack will be able to cause damage, while remaining undetected. Thus, our approach can be used to provide operators with real-time early warnings about stealthy attacks. 

\end{abstract}

\section{Introduction}\label{section:intro}

Industrial Control Systems (ICS) denote systems where safety-critical physical processes are augmented with computation and communication capabilities, e.g.\ transportation systems, manufacturing, and chemical processes. Recently, the security of ICS has received increasing attention, especially with the rise in the number of attacks against these systems, e.g.\ Ukrainian power grid blackout~\cite{Lee_Analysis_2016}. Differently from attacks targeting IT systems, attacks against ICS can also cause physical damage, rather than only harming digital assets, e.g. sensitive data. In particular, stealthy attacks, where resourceful attackers exploit noise~\cite{Bai_Security_2015} or control theoretic properties~\cite{Pasqualetti_Control_2015} to avoid detection, can cause significant damage. Although a variety of techniques~\cite{Giraldo_Survey_2018} consider the behaviour of the physical process to detect attacks on ICS, the detection of stealthy attacks still presents several limitations~\cite{Griffioen_Tutorial_2019}.
\par
Assessing the risk of stealthy attacks involves performing offline impact assessment~\cite{Murguia_reachable_2018,Hashemi_Gain_2019,Milosevic_Quantifying_2018,Urbina_Limiting_2016,Mo_performance_2016,Umsonst_Anomaly_2018}, which may provide operators with more insight into potential vulnerabilities, such as the inability of a residual-based anomaly detector to detect certain sensor attacks before they cause damage. However, offline impact assessment cannot account for potential transients and variations in operating modes that a physical system may experience. In particular, chemical plants often experience long transients and frequent changes in operating conditions due to potential unforeseen disturbances, real-time optimisation modules, or high-level control decisions~\cite{Mokhatab_Handbook_2012,Larsson_Self_2001}.  Safety analysis consists of checking whether, from a current state, the system can enter an unsafe state given the current control settings. Conventional control methods cannot guarantee the safety of the system given the possibility of stealthy attacks to exploit noise or control-theoretic properties to avoid detection~\cite{Kwon_Reachability_2018}. Therefore, there is a need for safety monitoring techniques that evaluate the safety of the system in real-time given such threats.
\par
The objective of this paper is to develop an efficient online safety monitoring algorithm, specifically under stealthy attacks on sensors. To the best of our knowledge, only a few works~\cite{Kwon_Reachability_2018,Coletta_Predictive_2018,Castellanos_modular_2019,Etigowni_Crystal_2018} fall into this line of research, and they either do not consider intelligently crafted stealthy attacks or are resource-intensive. Conversely, our approach to online safety monitoring provides a computationally-efficient online  mechanism to detect the potential impact of a range of stealthy attacks, that would be undectable using traditional monitoring approaches.
In terms of efficiency and scalability, the main feature of our approach is to perform the most computationally intensive operations offline, and reduce the online safety checks to a computation of a distance measure. Namely, the intensive offline computations consist of approximating symbolic reachable sets of states. When deployed online, these sets  only need to be instantiated depending on the current state estimate and a prediction of the state over a certain time horizon. We then take advantage of the geometric representation of the reachable sets  to perform efficient safety checks. This general approach is inspired by the work by Chen et al.~\cite{Chen_Model_2017} in the context of real-time monitoring for simplex control architectures.
\par
In this paper, we apply our approach to systems that can be approximated by a Linear Time-Invariant (LTI) model; we illustrate this model class using the large-scale Tennessee-Eastman Process (TEP). While we rely on previously established results on offline reachability analysis of these systems under stealthy attacks, our main contribution lies in applying these results to design an efficient and scalable online safety monitoring algorithm. 
We adapt the method originally developed by Murguia et al.~\cite{Murguia_reachable_2018} to compute reachable sets of estimation error under stealthy sensor attacks. In particular, our proposed algorithm first computes a symbolic ellipsoidal approximation of the reachable set offline, which is parametrised by the state estimate. Then, given a certain time horizon, it predicts the future states of the system, where for each state, we instantiate the reachable set under a potential attack. The algorithm terminates when it encounters a state which features a non-empty intersection with a predefined set of unsafe states, or when it exhausts the length of the horizon set for prediction. Since in most practical situations unsafe sets can be interpreted geometrically as a union of half-spaces~\cite{Murguia_Security_2018}, we use efficient ellipsoidal calculus techniques~\cite{boyd_convex_2004} to perform these emptiness checks.
\par
As a secondary contribution, we propose two online security metrics that can be computed by leveraging ellipsoidal calculus. The \emph{potential impact metric} quantifies the potential impact of a stealthy attack. When the emptiness check returns a negative result, the intersection between the reachable set and the set of unsafe states can be approximated using an ellipsoid. We use the size of this ellipsoid to quantify the potential impact. When the intersection between the reachable set and the set of unsafe states is non-empty, it is also possible to compute the \emph{time-to-unsafe metric}. This metric estimates the shortest time that an attacker would need to cause damage before being detected. This time-to-unsafe metric is fundamentally different from proximity-based metrics previously proposed in the literature~\cite{Castellanos_modular_2019,Carcano_multidimensional_2011}. These metrics rely on the raw estimate of the state of the system to compute an euclidean distance to unsafe states, and are used to perform safety monitoring. Instead, our time-to-unsafe metric relies on reachable sets induced by a potential stealthy attack. As such, we account for the fact that the given estimate may not represent the real state of the system.
\par
Finally, we  evaluate the proposed algorithm using the Tennessee-Eastman process (TEP) as a case study. We first validate our algorithm through extensive simulations aimed at assessing its ability to warn about potential damage due to a stealthy attack. Second, we compare it to existing online safety monitoring techniques for attacks, namely those that only rely on proximity-based metrics using raw state estimates. We show through simulation scenarios that under ``low-and-slow'' stealthy attacks, existing techniques will not convey the security and safety situation accurately. Conversely, our reliance on reachable sets in our approach allows for early warnings to be provided to operators before a stealthy attack can cause damage. Finally, we demonstrate the suitability of the algorithm to real-time applications. Specifically, we show that safety checking takes place in a time frame that is shorter than the system's sampling period, and that the algorithm scales well with the complexity of safety constraints and the desired length of time horizon for online prediction. We have applied our monitoring approach within a framework for physics-based early warnings for stealthy attacks~\cite{Azzam_Grounds_2021}.
\par
The rest of this paper is organised as follows: Section~\ref{section:related_work} discusses related work, Section~\ref{section:probstatement} provides an overview of our approach, Section~\ref{section:syslayout} describes the adopted modelling framework, Section~\ref{section:contribution} details the proposed algorithm, and Section~\ref{section:evaluation} presents numerical simulation results. Finally, Section VII concludes the paper.\
\section{Related Work}\label{section:related_work}
To the best of our knowledge, there is little work on the problem of online safety monitoring for ICS under stealthy attacks. Kwon et al.~\cite{Kwon_Reachability_2018,Kwon_Recursive_2016} have proposed a recursive method to compute exact reachable sets under stealthy attacks online. While this method is computationally efficient, it uses large recursive matrices, which can make an extensive use of resources. Furthermore, safety checking in this work relies on the characterisation of a time-varying safe set as an ellipsoid centred at the current state. Although this is suitable for the Unmanned Aerial Vehicle (UAV) application used by the authors, it may not be applicable in chemical process control, where unsafe operating levels are usually fixed limits imposed on physical state variables. In contrast, the bulk of the computation required for our method is performed offline, resulting in symbolic sets with a lightweight characterisation when instantiated online. Furthermore, we consider more practical time-invariant unsafe sets which can be interpreted geometrically as a union of half-spaces.
\par
Existing online monitoring schemes~\cite{Etigowni_Crystal_2018,Carcano_multidimensional_2011,Coletta_Predictive_2018,Castellanos_modular_2019} rely on a notion of proximity to a predefined set of unsafe/critical states. This line of work does not consider formal safety guarantees, but it uses metrics reflecting the proximity of the system to unsafe states as a way to either determine the level of safety or to detect attacks. For example, Carcano and Coletta~\cite{Carcano_multidimensional_2011,Coletta_Predictive_2018} compute the minimum Euclidean distance from current states to the unsafe operating region. Castellanos and Zhou~\cite{Castellanos_modular_2019} extend this notion further by computing an approximate ``time-to-critical-states'' metric. However, these approaches rely only on raw sensor values and do not consider the effect of stealthy attacks. For example, intelligently crafted sensor attacks introduce ``low-and-slow'' modifications to sensor values, which may eventually not reflect the real state of the system. In our work, instead of using raw sensor values, we rely on reachable sets under stealthy attacks which bound --- with a certain confidence level --- the actual state of the system.
\par
Other related work~\cite{Hashemi_Gain_2019,Milosevic_Quantifying_2018,Urbina_Limiting_2016,Mo_performance_2016,Umsonst_Anomaly_2018} has proposed techniques to quantify the worst-case impact of potential stealthy attacks. To the best of our knowledge, these techniques are developed with the objective of performing risk assessment offline. For example, Milosevic et al.~\cite{Milosevic_Security_2018a} propose a framework for security measure allocation given certain impact and attack complexity metrics. Murguia et al.~\cite{Murguia_Security_2018} use the volume of ellipsoidal approximations of reachable sets under stealthy attacks as a measure of impact. In our work, we quantify in real-time the potential impact of a stealthy attack based on the size of the intersection of the reachable set with the set of unsafe states. Using this intersection, instead of the entire reachable set, gives a more precise estimate of potential impact. This is made possible by using the geometric properties of the sets' representations.
\par
Finally, our approach is inspired by recent work on real-time reachability analysis~\cite{Althoff_Online_2015,Chen_Model_2017,Tran_Decentralized_2019}, notably in the context of real-time monitoring for simplex control architectures~\cite{Johnson_Real_2016}. Similarly to the algorithms proposed in this paper, a few works in this area consider pre-computing reachable sets offline before instantiating them online, in order to perform safety checking efficiently. However, to the best of our knowledge, previous work in this context did not consider safety checking in the presence of stealthy attacks seeking to cause damage to a safety-critical system.
\begin{figure*}[!t]
  \centering
  \includegraphics[width=\textwidth{},keepaspectratio]{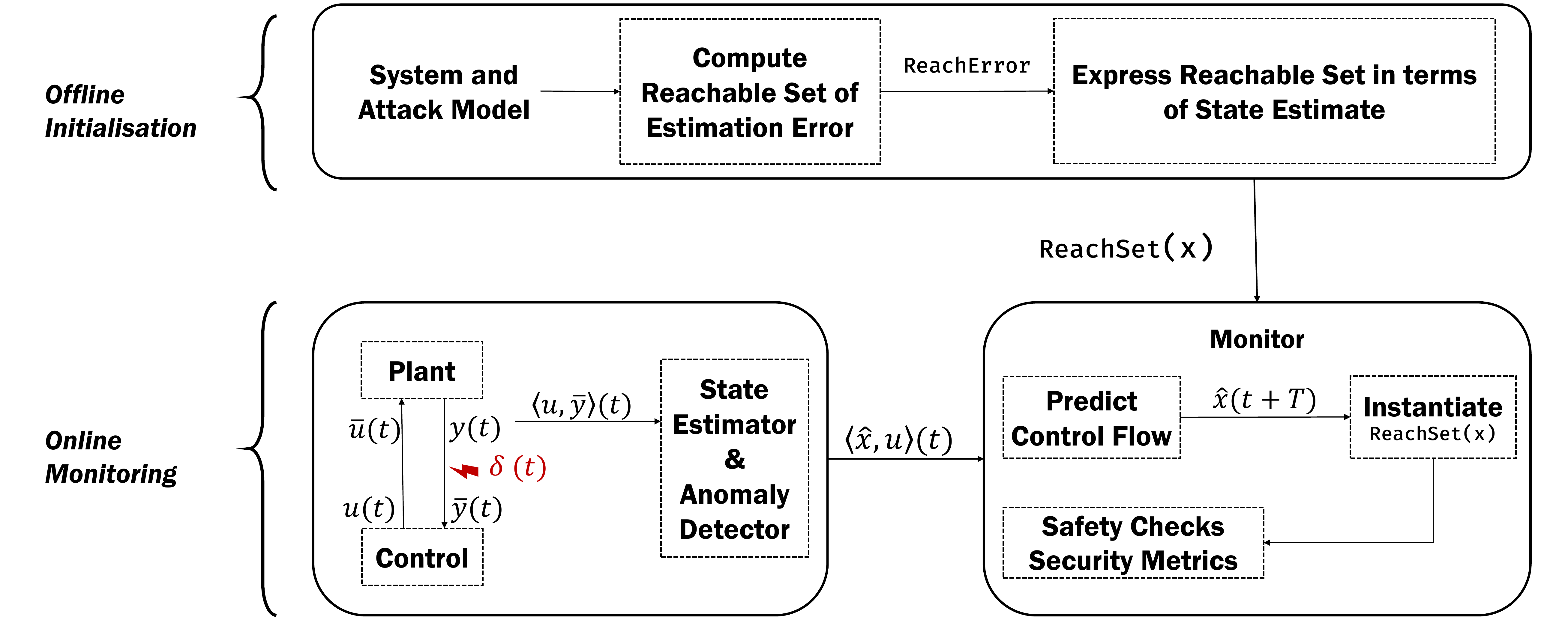}
  \caption{Outline of the proposed online monitoring approach.}\label{fig:scheme}
\end{figure*}
\section{Predictive Online Monitoring Approach}\label{section:probstatement}
In this section, we describe the online monitoring problem tackled in this paper and outline the proposed approach. We also describe the main idea behind existing work on online safety monitoring under attacks. We consider this body of work as a baseline against which we compare our approach.
\par
\subsection{Online Monitoring Problem}
This paper considers the online monitoring framework depicted in Figure~\ref{fig:scheme}. The objective of our approach is to check in real-time whether a potential undetected attack can cause damage to the system before being detected. In other words, given the current physical state estimate \(\Hat{x}(t)\), the online monitoring problem asks whether there exists a stealthy false data injection attack on sensors that can bring the system into an unsafe state over the next \(T\) time instants.
\par
If this check returns a negative result, then operators can be reassured that even if an intrusion is present, the alleged attacker may not be able to cause any damage without being detected. Otherwise, the check can serve as an early warning and can prompt operators to take preemptive safety or security measures. Such additional measures are however beyond the scope of the current work.
\par

\subsection{Outline of the Proposed Approach}
The proposed approach, shown in Figure~\ref{fig:scheme}, is composed of two main steps:
\begin{itemize}
  \item \textit{Offline Initialisation:} this step consists of computing symbolic reachable sets under a stealthy false data injection attack in terms of the state estimate based on a model of the system and attack. This is possible by considering the evolution of the state estimation error under a stealthy attack instead of the physical state itself. As a result, we express the reachable set of errors in terms of the actual state estimate. This allows us to perform the bulk of the computation offline, leading to more efficient real-time safety checking.
  \item \textit{Online Monitoring:} at runtime, the proposed monitor takes as input the current physical state estimate and the state of the controller and predicts the value of the state up to \(T\) time steps into the future. We assume that this prediction can be done using an identified physical model of the system, with \(T\) chosen to maintain an acceptable degradation in the confidence level of the predicted state. We then instantiate the precomputed symbolic reachable set at each predicted state value, and perform an emptiness check of its intersection with a predefined set of unsafe states. The prediction stops when a non-empty intersection is encountered, or when \(T\) is exhausted. Furthermore, we compute two security metrics when the intersection is non-empty: (i) \textit{potential impact} of the attack, and (ii) \textit{time-to-unsafe states} reflecting the approximate time that a potential attack would need to cause damage. These metrics are aimed at providing operators with a better assessment of the current safety/security situation.
\end{itemize}
\par
In this paper, we apply this general approach to LTI systems, where we propose the use of ellipsoids as over-approximations of the symbolic reachable sets. Ellipsoids have been extensively used for safety verification for control systems~\cite{kurzhanski_ellipsoidal_2000,hu_reach_2020}. They feature an efficient quadratic representation in terms of the dimension of the state of the system~\cite{leguernic_reachability_2009}, which presents an advantage in real-time monitoring. Furthermore, in most practical applications of process control, unsafe operating regions can be represented as unions of half-spaces. With reachable sets represented as ellipsoids, safety checking reduces to checking the sign of the distance from the ellipsoid to each of the hyperplanes composing the unsafe set~\cite{boyd_convex_2004,kurzhanski_ellipsoidal_2000}. As a result, real-time safety monitoring is enabled with minimal resource utilisation.
\par
In our application to LTI systems, we use results by Murguia~\cite{Murguia_reachable_2018} in reachability analysis under stealthy attacks to precompute a symbolic reachable set in the offline initialisation phase. We then use results in ellipsoidal calculus~\cite{kurzhanski_ellipsoidal_2000,boyd_convex_2004} to design an efficient and scalable online safety monitoring algorithm for stealthy attacks.

\subsection{Existing Monitoring Techniques}\label{subsection:existing_monitoring_tech}
To the best of our knowledge, existing online safety monitoring techniques~\cite{Etigowni_Crystal_2018,Carcano_multidimensional_2011,Coletta_Predictive_2018,Castellanos_modular_2019} rely mainly on a proximity metric to assess the safety of the system against attacks targeting physical processes. In this paper, we compare our approach to these techniques which all feature the same idea detailed in the following. We provide a comparison based on intuition in this section.
\par
Given a set of unsafe states \(\mathcal{S}_u\), and the current estimated state \(\Hat{x}(t)\), existing online monitoring techniques~\cite{Etigowni_Crystal_2018,Carcano_multidimensional_2011,Coletta_Predictive_2018,Castellanos_modular_2019} compute a distance metric \(d_u = \min{d(\Hat{x}(t),\mathcal{S}_u)}\). The most commonly used distance metric is the Euclidean distance, which is mainly suitable for continuous or hybrid systems where state variables of interest assume real values. Given a set of thresholds \(\tau_1 > \tau_2 > \tau_3 \dots\), alarms of different levels of criticality can be raised based on the value of \(d_u\). The proximity metric can further account for the dynamics of the system by computing a \textit{time-to-unsafe/critical states} metric, \(t_u = d_u/r\); where \(r\) is the approximate rate of change of the system state over a given time period~\cite{Castellanos_modular_2019}. In the rest of the paper, the term ``traditional time-to-unsafe metric'' refers to \(t_u\) computed using the aforementioned formula.
\par
However, under a stealthy attack which slowly drives the system into unsafe states to avoid detection, the real state of the system \(x(t)\) will diverge from the estimated state. Thus, the metric \(d_u\) may not provide an accurate measure of the proximity of the system to unsafe states. Instead of relying merely on a proximity measure based on \(\Hat{x}(t)\), our online monitoring approach accounts for the possibility of stealthy attacks by considering reachable sets under such attacks. Namely, if \(\mathcal{R}_x(t)\) is the reachable set of states under a stealthy attack given the used anomaly detector at time \(t\), then \(x(t) \in \mathcal{R}_x(t)\) even if \(x(t)\neq \Hat{x}(t)\). Otherwise, by definition, the attack would be detected by the anomaly detector. If over the next \(T\) time period, \(\mathcal{R}_x(t') \cap \mathcal{S}_u \neq \emptyset\) for some \(t'\in [t;t+T]\), then it is possible for an attacker to drive the system into the unsafe operating region within at least \(t' - t\) time units. As such, our safety monitoring approach relies on emptiness checking of this intersection, instead of a mere proximity measure. Furthermore, we consider the proximity metric \(t'-t\) as an additional security metric that may assist operators in assessing the current safety/security situation.
\par

\section{Modelling Framework}\label{section:syslayout}
We apply our approach in this paper to Linear Time-Invariant (LTI) systems. Before detailing the proposed approach, we briefly describe the layout of the system and the attack model in this section.
\par
\subsection{System Layout}
Consider the control system architecture in Figure~\ref{fig:scheme}.\ We assume that the physical system can be approximated using a Linear Time-Invariant (LTI) model:
\begin{equation}\label{eq:sys.ltimodel}
    \left \{
      \begin{aligned}
        x(k+1) &= Ax(k) + B\Bar{u}(k) + w(k)\\
        y(k) &= Cx(k) + v(k)
      \end{aligned}
    \right.
  \end{equation}
Where \(x(k)\in\R^n\) is the state vector, \(\Bar{u}(k)\in \R^l\) denotes control signals received by the system, and \(y(k)\in\R^m\) is a vector of sensor measurements. \(w(k)\in\R^n\) and \(v(k)\in\R^m\) denote process and measurement noise, respectively, and are assumed to follow a zero-mean Gaussian distribution with respective covariance matrices \(\Sigma_1\) and \(\Sigma_2\). \(k = t/\Delta_t \in \mathbb{N}\) denotes discrete time instants, where \(\Delta_t\) is the sampling period and \(t\) is continuous time. \(A,\; B,\;\text{and}\; C\) are real, time-invariant matrices of appropriate dimensions. The system is assumed to be equipped with an output feedback control loop, such that \(u(k) = \mathcal{K}(\Bar{y} - y_r(k))\), where \(u(k)\) denotes control signals originally sent to the process, \(\mathcal{K}\) is the control law, \(y_r(k)\) denotes reference output, and \(\Bar{y}(k)\) denotes measurements received by the controller. In this work, we focus on sensor attacks, such that \(u(k) = \Bar{u}(k)\;\forall k\).
\par
Furthermore, we assume that a subset of state variables, denoted as critical, are grouped in the vector \(x_c = C_c x,\; x_c\in\R^{n_c},\; C_c \in \R^{n_c\times n}\), and are required to remain within a certain safe set to ensure safe operation. Let \(\mathcal{S}_u\) be the set of unsafe states; we assume that unsafe conditions are given as a linear combination of the critical state variables, such that the unsafe set becomes a union of half-spaces\footnote{This assumption is typical in several process control applications.}:
\begin{equation}\label{eq:sys.unsafeset}
    \mathcal{S}_u = \left \{x(k)\in\R^n\;|\; \bigcup_{i=0}^{n_c} C_{c,i} x_{i}(k) \geq b_i \right \}
\end{equation}
Where \(b_i\in\R \) is the \(i^{\text{th}}\) half-space scalar, \(C_{c,i}\) denotes the \(i^{\text{th}}\) row of the matrix \(C_c\), and \(x_i(k)\) the \(i^{\text{th}}\) element of \(x(k)\).
\par
At a time \(k\), given previous sensor measurements and control actions, a Kalman filter generates an estimate of the physical state and expected sensor measurements as follows:
\begin{equation}\label{eq:sys.kalman}
    \left \{
    \begin{aligned}
        \Hat{x}(k) =& \; A\Hat{x}(k-1) + Bu(k-1) 
                 \\ &+ L(\Bar{y}(k-1) - C\Hat{x}(k-1))\\
        \Hat{y}(k) =& \; C\Hat{x}(k)
    \end{aligned}
    \right.
\end{equation}
Where \(L\) denotes the Kalman gain, and \(\Hat{x}\) and \(\Hat{y}\) denote estimated state and measurements, respectively.
\par
In addition, a chi-squared anomaly detector compares received measurements with the generated estimate through a residual \(r(k):=\Bar{y}(k)-\Hat{y}(k)\). Under nominal conditions, the residual metric has a zero-mean and a covariance matrix \(\Sigma \). To check for this hypothesis, a chi-squared metric, \(z(k) = r^T(k)\Sigma^{-1}r(k)\) is computed and compared with a threshold \(\tau \), such that exceeding this threshold implies a possible anomaly and raises an alarm. The threshold \(\tau \) is designed to maintain a certain false alarm rate \(\beta\) such that \(\Pr[z(k) \leq \beta] = 1-\beta\) under nominal operation, and can be set as described by Murgia et al.~\cite{Murguia_reachable_2018} (Proposition 1).

\subsection{Attacker Model}
We consider in this work false data injection attacks on sensors which are masked by the system noise, in order to drive the latter slowly to the unsafe set (\ref{eq:sys.unsafeset}). We assume that the attacker has sufficient resources and knowledge about the system to carry out this attack. Let \( \{k_s,\dots, k_f \} \) denote the time period of the attack; we model the attack as a bias imposed on sensor measurements:
\begin{equation}\label{eq:attack}
    \Bar{y}(k) := y(k) + \delta(k) \; \forall \; k\in \{k_s,\dots, k_f \}
\end{equation}
The attack \(\delta(k) \) remains stealthy by ensuring that the false alarm rate is maintained throughout the attack period. We use this characterisation because anomaly detectors do raise alarms under nominal operation. A sudden disappearance of these alarms in practice may raise suspicion in operators and lead them to uncover the attack before it can cause damage~\cite{Hashemi_comparison_2018}. Under the attack in (\ref{eq:attack}), the residual is given by: \(r(k) = \Bar{y}(k) - \Hat{y}(k) = y(k) - \Hat{y}{k} + \delta(k) \). As such, the chi-squared metric under attack is given by:
\begin{equation}\label{eq:detectmetric}
    z(k) = (y(k) - \Hat{y}(k) + \delta(k))\Sigma^{-1}(y(k) - \Hat{y}(k) + \delta(k))
\end{equation}
By selecting \(\delta(k) \) to be such that \( \Pr[z(k) \leq \tau] = 1 - \beta \), the attacker manages to remain undetected. For example, given \(K\) time steps, the attacker may choose to raise alarms for \(\beta K\) steps so that the false alarm rate is mimicked as closely as possible~\cite{Hashemi_comparison_2018}\footnote{Due to space limitations, the reader is referred to~\cite{Hashemi_comparison_2018} for a more detailed description of the distribution of the detector metric under such attack.}. This is possible since we assume that the attacker knows the detector's parameters, i.e. \(\Sigma \), \(\beta\) and \(\tau \), in addition to the system and estimator outputs; i.e. \(y(k)\) and \(\Hat{y}(k)\) respectively. In practice, such information could be obtained, for example, through reconnaissance attacks or insider knowledge. 
\section{Proposed Monitoring Algorithm}\label{section:contribution}
The proposed approach relies mainly on the offline computation of the symbolic reachable set of estimation error under the attack described in (\ref{eq:attack}). This set is an ellipsoidal over-approximation of the exact reachable set, parametrised by the state estimate. In real-time, given the state estimate at time \(k\), the symbolic set is instantiated at the \(K\)-step predicted state. Emptiness checks of its intersection with \(\mathcal{S}_u\) are then performed. We detail both the offline and online computations in the following sections, and we propose online security metrics based on the computed reachable set.

\subsection{Offline Computation of Symbolic Reachable Set}
To compute the reachable set of the estimation error under the attack in (\ref{eq:attack}), we use the method described by Murgia et al.~\cite{Murguia_reachable_2018}. We define this error as \(e(k):=x(k) - \Hat{x}(k)\), and assume that at the start of an attack the estimation error is always almost zero. The reachable set of the estimation error under the attack in (\ref{eq:attack}) is independent of the actual physical state at the start of the attack. As such, this set serves as a symbolic reachable set parametrised by the state estimate.
\par
By setting \(e(k) = x(k)-\Hat{x}(k)\), and performing some algebraic manipulations of Equation (\ref{eq:sys.kalman}), the evolution of the estimation error under an attack is given by:
\begin{equation}\label{eq:estimationerror}
    \begin{aligned}
    e(k+1) &= Ae(k) - L(Ce(k) + v(k) + \delta(k)) + w(k)\\
           &= Ae(k) - L(y(k) - \Hat{y}(k) + \delta(k)) + w(k)
    \end{aligned}
\end{equation}
\par
Since the error is partially driven by the Gaussian noise \(w(k)\) and the attack-dependent sequence \(\Bar{\delta}(k) = y(k) - \Hat{y}(k) + \delta(k) \), using a deterministic approach will yield an unbounded reachable set, as the support of \(w(k)\) and \(\delta(k)\) (as characterised in (\ref{eq:attack}) and (\ref{eq:detectmetric})) is infinite. This issue can be overcome by setting a confidence level on the energy of both of these vectors. For the attack, the sequence \(\Bar{\delta}(k)\) is already constrained to be such that \(\Pr[z(k)\leq \tau] = \Pr[\lVert \Sigma^{-1/2}\Bar{\delta}(k) \rVert^2\leq \tau]= 1-\beta \) where \(\lVert . \rVert \) denotes the \(L_2\)-norm. For the noise, let \(p = \Pr[\lVert w(k)\rVert^2 \leq \Bar{w}]\); since \(w(k)\) follows a zero-mean Gaussian distribution, the bound \(\Bar{w}\) on \(\lVert w(k)\rVert^2\) can be determined using the gamma distribution for a desired confidence \(p\).
\par
By using this assumption, the resulting reachable set can be interpreted as a level set of the distribution of the reachable error. A larger confidence level would lead to a larger set, at the cost of being overly conservative with the safety checking. A reasonable choice for \(p\) would be \(1-\beta \), as the false alarm \(\beta \) is designed to be small. This also simplifies the computation of the reachable set, since for \(p=1-\beta \), we readily have \(\Pr[\lVert w(k)\rVert^2 \leq \Bar{w}] = \Pr[z(k)\leq\tau]\) under the attack in (\ref{eq:attack}). The following is based on this choice; for a more detailed treatment of this confidence level and a comparison of reachable sets under different choices of \(p\), the reader is referred to~\cite{Murguia_reachable_2018} and~\cite{Hashemi_comparison_2018} due to space limitations. Let \(\mathcal{R}_e^p\) denote the reachable set of error under the attack in (\ref{eq:attack}) and a confidence level \(p = 1-\beta \):
\begin{equation}
    \begin{aligned}
    \mathcal{R}_e^p := \{e(k)\in\R^n\;|\; &e(k)\; \text{is s.t. (\ref{eq:estimationerror})},\\ p&=\Pr[\lVert w(k)\rVert^2 \leq \Bar{w}] = 1 - \beta \}
    \end{aligned}
\end{equation}
While computing \(\mathcal{R}_e^p\) is intractable, it is possible to over-approximate the set using an ellipsoid in \(\R^n\), given by:
\begin{equation}
    \mathcal{R}_e^p \subseteq \mathcal{E}_e^p = \{e(k)\;|\; e^T(k)\mathbf{\Pi}^{-1}e(k) \leq 1\}
\end{equation}
Where the positive definite matrix \(\mathbf{\Pi}\) is the ellipsoid's shape matrix. Letting \(\mathcal{P} = \PI^{-1}\), the minimum volume ellipsoid containing the set \(\mathcal{R}_e^p\) can be obtained by solving the following semi-definite programme~\cite{Murguia_reachable_2018}:
\begin{equation}\label{eq:shapematrix}
    \begin{aligned}
        \mathcal{P} =& \argmin -\log\det \mathbf{\mathcal{P}} \\
        &\text{s.t.}\\
        &\mathcal{P} > 0 \; ;\; \mathcal{Q} \geq \mathbf{0}
    \end{aligned}
\end{equation}
Where:
\begin{equation}
    \begin{aligned}
    \mathcal{Q} &= 
    \begin{bmatrix}
        b\mathcal{P} & A^T\mathcal{P} & \mathbf{0} & \mathbf{0} \\
        \mathcal{P} A & \mathcal{P} & \mathcal{P} & -\mathcal{P} L \Sigma^{1/2} \\
        \mathbf{0} & \mathcal{P} & \frac{1-b}{\tau+\bar{w}}I & \mathbf{0} \\
        \mathbf{0} & -\Sigma^{1/2}L^T\mathcal{P} & \mathbf{0} & \frac{1-b}{\tau+\bar{w}}I \\ 
    \end{bmatrix}
    ,\\ b &\in (0,1)
    \end{aligned}
\end{equation}
Note that while \(b\) is an optimisation variable, it is necessary to fix it to ensure the convexity of the programme. A grid search can then be performed over the interval \((0,1)\) to find the optimal shape matrix corresponding to the minimum-volume ellipsoid.
\par
Given the shape matrix \(\PI = \mathcal{P}^{-1} \), and replacing \(e(k)\) by its definition,
we obtain a symbolic ellipsoidal approximation \(\mathcal{E}_x^p(k)\) of the reachable set \(\mathcal{R}_x^p(k)\) of the actual system state \(x(k)\), parametrised by the current state estimate \(\Hat{x}(k)\):
\begin{equation}\label{eq:reachsetx}
    \begin{aligned}
    &\mathcal{R}_x^p(k) \subseteq \mathcal{E}_x^p(k) =\\ 
    &\{x(k)\in\R^n\;|\;{(x(k)-\Hat{x}(k))}^T\PI^{-1}(x(k)-\Hat{x}(k)) \leq 1\}
    \end{aligned}
\end{equation}
\par
\begin{algorithm}[!t]
    \caption{Offline Symbolic Reachable Set Computation}\label{alg:offline}
    \begin{algorithmic}[1]
    \Statex{}\textsc{Inputs:} (\(A,L,\Sigma,\tau,\Bar{w},\Delta h\)); \(0 < \Delta h < 1\)
    \Statex{}\textsc{Output:} Reachable set shape matrix \(\PI\)
    \Statex{}
    \State{}\(b \gets \Delta h\);
    \State{}SolutionList \(\gets \)\textit{EmptyList}();
    \While{\(b < 1\)}
        \Statex{}\Comment{Solve the programme in (\ref{eq:shapematrix}) for the current value of \(b\)}
        \State{}\textit{SolveSemiDefiniteProgramme}(\(A,L,\Sigma,\tau,\Bar{w},b\));
        \State{}SolutionList.\textit{append}(CurrentSolution);
        \State{}\(b \gets b+\Delta h\);
    \EndWhile{}
    \State{}BestShapeMatrix \(\gets \) \textit{MinObjectiveValue}(SolutionList);
    \State{}\textbf{return} BestShapeMatrix;
    \end{algorithmic}
\end{algorithm}
Algorithm~\ref{alg:offline} summarizes the offline steps to obtain \(\mathcal{E}_x^p(k)\). Given the system matrix \(A\), the Kalman gain \(L\), the residual covariance matrix \(\Sigma \), the anomaly detector's threshold \(\tau \) and the confidence bound \(\Bar{w}\), the algorithm performs a grid search over \((0,1)\) by partitioning the interval  into segments of length \(\Delta h\). The choice of \(\Delta h\) will depend on the desired tightness of the ellipsoidal approximation given the computational resources available. Note that this step only needs to be performed offline once, and only the matrix \(\PI \) needs to be stored to instantiate \(\mathcal{E}_x^p(k)\) online given a state estimate \(\Hat{x}(k)\).

\subsection{Online Safety Checks}
\begin{algorithm}[!t]
    \caption{Online Safety Checking}\label{alg:online}
    \begin{algorithmic}[1]
    \Statex{}\textsc{Inputs:} (\(K, \PI ,\Hat{x}(k)\), ControllerState, UnsafeSet)
    \Statex{}\textsc{Output:} \textit{true} if the system is safe under a potential stealthy attack; \textit{false} otherwise 
    \Statex{}
    \State{}\(\Hat{x}_p \gets \Hat{x}(k)\)
    \ForAll{\(l\in \{0,1,\dots,K\}\)}
        \State{}ReachEll \(\gets \)\textit{Ellipsoid}(\(\Hat{x}_p\),\(\PI \));
        \ForAll{Hyperplane \(\subset \) UnsafeSet}
            \State{}DistToUnsafe \(\gets \) \textit{dist}(ReachEll,Hyperplane);
            \If{DistToUnsafe \(\leq 0\)}
                \State{}\textbf{return} \textit{false};
            \EndIf{}
        \EndFor{}
        \State{}\(\Hat{x}_p\gets\) \textit{PredictControlFlow}(\(\Hat{x}_p\),ControlState);
    \EndFor{}
    \State{}\textbf{return} \textit{true};
    \end{algorithmic}
\end{algorithm}
Algorithm~\ref{alg:online} outlines the steps needed to perform online safety checks. Given the current state estimate \(\Hat{x}(k)\) and the state of the controller, we estimate the state of the system for \(K\) time steps into the future using the identified model of the plant. At each time step \(l\in\{0,\dots,K\}\), we instantiate \(\mathcal{E}_x^p(k+l)\) and we check whether it intersects the set \(\mathcal{S}_u\). The algorithm halts and reports an unsafe state when a non-empty intersection is encountered. If the prediction horizon is exhausted, the algorithm reports a safe state. In the following, we detail the procedure we use to perform the emptiness checks.
\par
Let \(\mathcal{H}_i = \{ x\in\R^n\;|\; C_{c,i} x \geq b_i \} \) be a half-space representing one of the safety conditions composing the set \(\mathcal{S}_u\) (Equation (\ref{eq:sys.unsafeset})). Checking whether \(\mathcal{E}_x^p(k_f)\cap\mathcal{S}_u = \emptyset \) involves checking whether \(\mathcal{E}_x^p(k_f)\cap\mathcal{H}_i = \emptyset \) for each \(i\in \{1,\dots,n_c\} \). If the latter is true for all \(i\), then the former is also true, since \(\mathcal{S}_u = \cup_{i=1}^{n_s} \mathcal{H}_i\). 
\par
To check whether \(\mathcal{E}_x^p(k_f)\cap\mathcal{H}_i = \emptyset \), it suffices to compute the minimum distance from \(\mathcal{E}_x^p(k_f)\) to the hyperplane that delimits the half-space \(\mathcal{H}_i\). Let \(\mathcal{H}_{p,i} = \{x\;|\; C_{c,i} x = b_i\} \) be such hyperplane, the minimum distance from \(\mathcal{E}_x^p(k_f)\) to \(\mathcal{H}_{p,i}\) is given by~\cite{kurzhanskiy_ellipsoidal_2006}:
\begin{equation}
    d_i(k_f) = \frac{\rvert b_i - C_{c,i} x(k_f)\lvert - \sqrt{{x(k_f)}^T\PI x(k_f)}}{\lVert C_{c,i}^T \rVert}
\end{equation}
If \(d_i(k_f) \leq 0\), then \(\mathcal{E}_x^p(k_f)\cap\mathcal{H}_i \neq \emptyset \). Otherwise, if \( d_i(k_f) \geq 0\), then the ellipsoid \(\mathcal{E}_x^p(k_f)\) is either contained in \(\mathcal{H}_i\) or does not intersect the half-space, depending on whether its centre \(\Hat{x}(k_f)\) belongs to \(\mathcal{H}_i\). However, since the state estimate is within the safe region\footnote{Otherwise, it would be clear that the system is evolving to an unsafe state and Algorithm~\ref{alg:online} in this case would become obsolete.}, i.e. \(\Hat{x}(k_f) \notin \mathcal{H}_i\), then in our case, \( d_i(k_f) > 0\) always implies that \(\mathcal{E}_x^p(k_f)\cap\mathcal{H}_i = \emptyset \).

\subsection{Real-time Security Metrics}
In addition to checking the emptiness of the intersection of the reachable set with the set of unsafe states, it is possible to derive two online security metrics. The first metric can help operators get a better idea of the potential impact of a stealthy false data-injection attack, while the second approximates minimum amount of time that would be required for an attacker to cause damage. In this section we show how ellipsoidal methods can be used to compute such metrics efficiently.

\subsubsection{Real-Time Impact of Stealthy Attack}
In the case where \(\mathcal{E}_x^p(k_f)\cap\mathcal{S}_u \neq \emptyset \), we can quantify the impact of a potential stealthy false data-injection attack using the approximate size of this intersection. Namely, for each half-space \(\mathcal{H}_i \subset \mathcal{S}_u\) it is possible to over-approximate \(\mathcal{E}_x^p(k_f)\cap\mathcal{H}_i \) using a minimum-volume ellipsoid \(\mathcal{E}_i(k_f)\) of centre \(q_i(k_f)\in\R^n\) and shape matrix \(\PI_i(k_f)\), as follows~\cite{boyd_convex_2004}:
\begin{equation}\label{eq:intersection}
    \begin{aligned}
        q_i(k_f) &= \Hat{x}(k_f) - \frac{1+\alpha_i n}{n+1}\PI\Bar{c}_i
        \\
        \PI_i(k_f) &= \frac{n^2(1-\alpha_i^2)}{n^2-1}\times \\
        &\left(\PI - \frac{2(1+\alpha_i n)}{(n+1)(\alpha_i + 1)}\PI\Bar{c}_i\Bar{c}_i^T\PI \right) 
    \end{aligned}
\end{equation}
Where \(\Bar{c}_i = C_{c,i}/{(C_{c,i}\PI C_{c,i}^T)}^{0.5}\) and \(\alpha_i = (C_{c,i}\Hat{x}(k_f)-b_i)/{(C_{c,i}\PI C_{c,i}^T)}^{0.5}\). As such, we quantify the impact of a potential stealthy false data-injection attack using the volume of \({\mathcal{E}_i(k_f)} \). The volume of a general ellipsoid in \(\R^n\) with a shape matrix \(Q\) is given by:
\begin{equation}\label{eq:impact-stealthy-attack}
    \mathbf{vol}(\mathcal{E}) = \mathbf{vol}[\mathcal{B}_n] \sqrt{\det{Q}}
\end{equation}
Where \(\mathbf{vol}[\mathcal{B}_n]\) and \(\det{Q}\) denote the volume of the unit \(n\)-ball and the determinant of the matrix \(Q\), respectively. It is worthwhile to note that different system dimensions may lead to vastly different number ranges for the volume of the intersection ellipsoid. Thus, in order to make the impact metric more meaningful, we propose to use the ratio of the volume of the intersection ellipsoid to that of the ellipsoid approximating the reachable set. This guarantees that the impact metric will fall in the range \([0;1]\), thus becoming more intuitive to interpret. From (\ref{eq:impact-stealthy-attack}), the impact metric reduces to the following:
\begin{equation}
    \mathbf{Im}(k) = [\max_{i=1\dots n_c}{\det{\PI_i(k_f)}}]/\det{\PI}
\end{equation}

\subsubsection{Approximate Time to Unsafe States}
In the case \(\mathcal{E}_x^p(k_f)\cap\mathcal{S}_u \neq \emptyset \) for some \(k_f\in\{k,\dots,k+K\}\), we use the time \(k_f\) as the approximate time to unsafe states metric in our approach, namely:
\begin{multline}
    \mathbf{Tc}(k) = (k_f - k)\Delta_t,\;\\ \exists k_f\in\{k,\dots,k+K\}\;\text{s.t.}\;\mathcal{E}_x^p(k_f)\cap\mathcal{S}_u \neq \emptyset
\end{multline}
\par
Where \(\Delta_t\) is the system's sampling period. The advantage of using \(k_f\) instead of the distance from the state estimate \(\Hat{x}(k)\) itself is that the former approach accounts for the fact that if an undetected attack is present, then the actual state \(x(k_f) \neq \Hat{x}(k_f)\) still lies within \(\mathcal{E}_x^p(k_f)\) (otherwise, the attack would be detected). This metric provides operators with an idea of the minimum time they have to react before a potential stealthy false data injection attack manages to bring the system into an unsafe state.

\section{Evaluation}\label{section:evaluation}
In this paper, we use the Tennessee-Eastman Process (TEP) with the control architecture designed by Ricker et al.~\cite{Ricker_Decentralized_1996} as a case study. We implemented Algorithms 1 and 2 in MATLAB, with the ellipsoidal techniques based on the Ellipsoidal Toolbox written by Kurzhanskiy et al.~\cite{kurzhanskiy_ellipsoidal_2006}. We performed our evaluation using a simulation of the TEP provided by Bathelt et al.~\cite{Bathelt_2015_Revision} and implemented in Simulink. We approximated the TEP process as an LTI system with 50 state variables using MATLAB's \texttt{n4sid} algorithm. We considered the safety constraints discussed by Down and Vogel~\cite{Downs_plant_1993}, shown here in Table~\ref{table:safety_constraints}. We augmented the TEP simulation with a Kalman filter and a chi-squared anomaly detector. To initialise the online monitoring tool, we ran Algorithm~\ref{alg:offline} to determine the reachable ellipsoid's shape matrix. We performed the grid search for parameter \(b\) by dividing the interval into segments of length \(0.01\).
\begin{table}[!t]
    \renewcommand{\arraystretch}{1.3}
    \caption{Safety constraints considered for the TE case study~\cite{Downs_plant_1993}.}\label{table:safety_constraints}
        \begin{center}
        \begin{tabular}{ c||c||c }
        \hline
        \textbf{Output} & \textbf{Low Limit} & \textbf{High Limit}\\
        \hline \hline
        Reactor Pressure & none & 2895 kPa\\
        \hline
        Reactor Temperature & none & 150 \(^{\circ} C\) \\
        \hline
        Reactor Level & 11.8 \(m^3\) & 21.3 \(m^3\) \\
        \hline
        Product Separator Level & 3.3 \(m^3\) & 9.0 \(m^3\) \\
        \hline
        Stripper Base Level & 3.5 \(m^3\) & 6.6 \(m^3\) \\
        \hline
        \end{tabular}
    \end{center}
\end{table}
\par
Our evaluation consists of three main parts. First, we validated our approach by measuring true and false positives rates using extensive simulations. Second, we compared our approach with existing online monitoring approaches. Finally, we assessed the performance and scalability of our approach.
\par
\subsection{Validation}
The objective of our approach is not to detect attacks, but rather to perform safety checking under \textit{potential} stealthy attacks that seek to cause damage. Namely, Algorithm 2 checks whether the current state of the system can be taken to an unsafe state by a stealthy attack within the next \(K\) time instants and cause damage before the anomaly detector detects the attack. Thus, to evaluate our approach, we ran several simulations of the TEP only considering different stealthy attacks on the sensors that report values of safety critical parameters shown in Table~\ref{table:safety_constraints}. We avoided using the true/false positive/negative rate  performance metrics as traditionally defined in the attack detection literature. Instead, we considered a notion of true/false positives/negatives similar to the one adopted in previous work on online safety monitoring~\cite{Chou_Predictive_2020}:
\begin{itemize}
    \item A \textit{true positive} occurs when Algorithm 2 raises a warning within \(K\) time instants before damage  occurs due to an attack, and the system reaches an unsafe state before the anomaly detector raises an alarm. 
    \item A \textit{true negative} occurs when Algorithm 2 does not raise any warnings within \(K\) time instants before damage  occurs, but the attack is detected by the anomaly detector. 
    \item A \textit{false positive} occurs when Algorithm 2 reports an unsafe state within \(K\) time instants before damage  occurs, but the anomaly detector manages to detect the attack before the system enters the unsafe state. 
    \item A \textit{false negative} occurs when Algorithm 2 does not raise any warnings within  \(K\) time instants before damage occurs, even if a stealthy attack is taking place, and the anomaly detector does not raise any alarm.
\end{itemize}
\par
\begin{figure}[!t]
    \centering
    \includegraphics[width = \columnwidth, keepaspectratio, angle = 0]{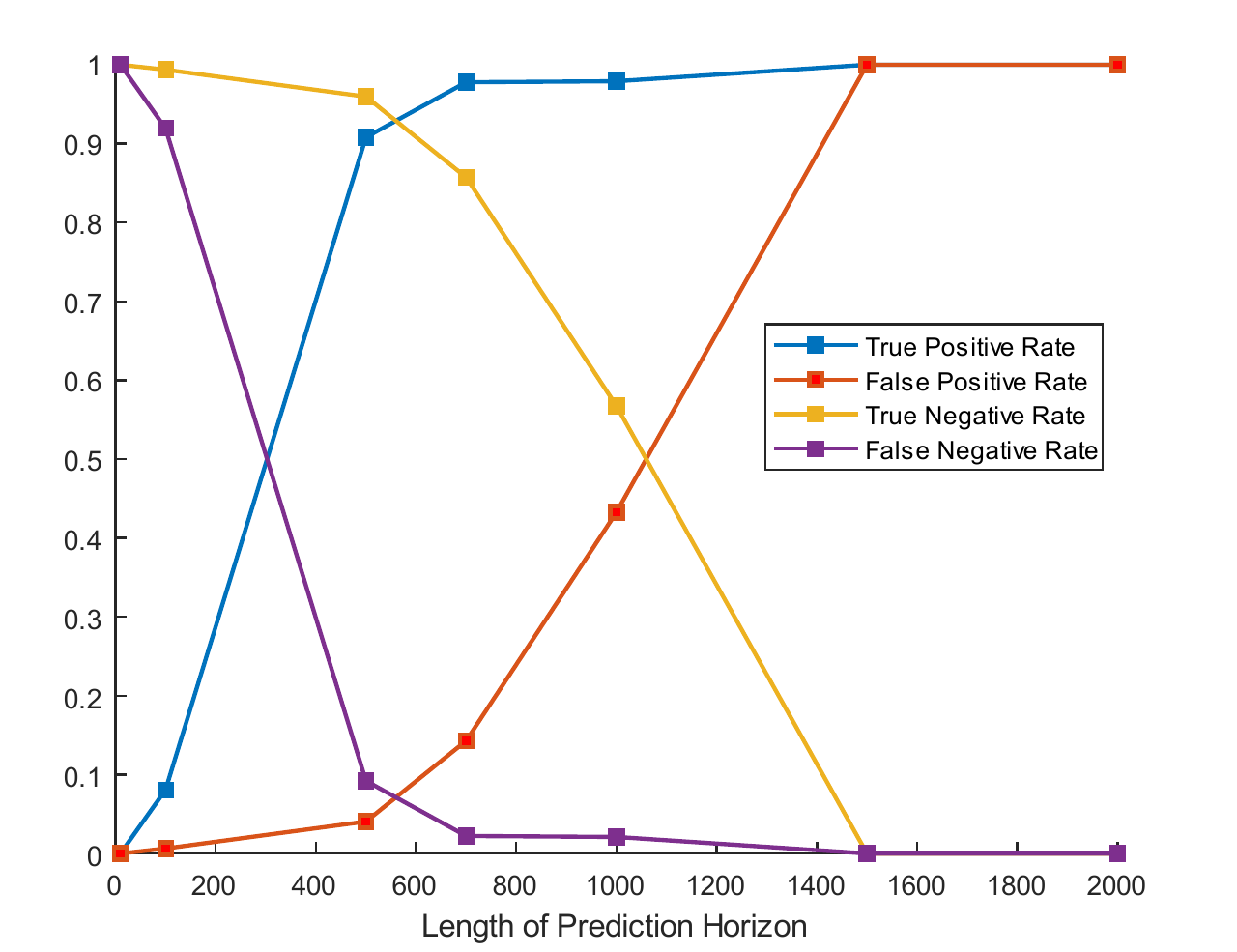}
    \caption{True/false positive/negative rates as a function of the length of prediction horizon \(K\).}\label{fig:results_validity_numsteps}
\end{figure}

 We first tested the effect of the length of the prediction horizon \(K\) on these rates, with results presented in Figure~\ref{fig:results_validity_numsteps}. For each value of \(K\) that we tested, we ran \(500\) simulations where we picked the attacked sensors at random, and we simulated the attack as a slowly growing bias on sensor measurements.
\begin{figure}[!t]
    \centering
    \includegraphics[width = \columnwidth, keepaspectratio, angle = 0]{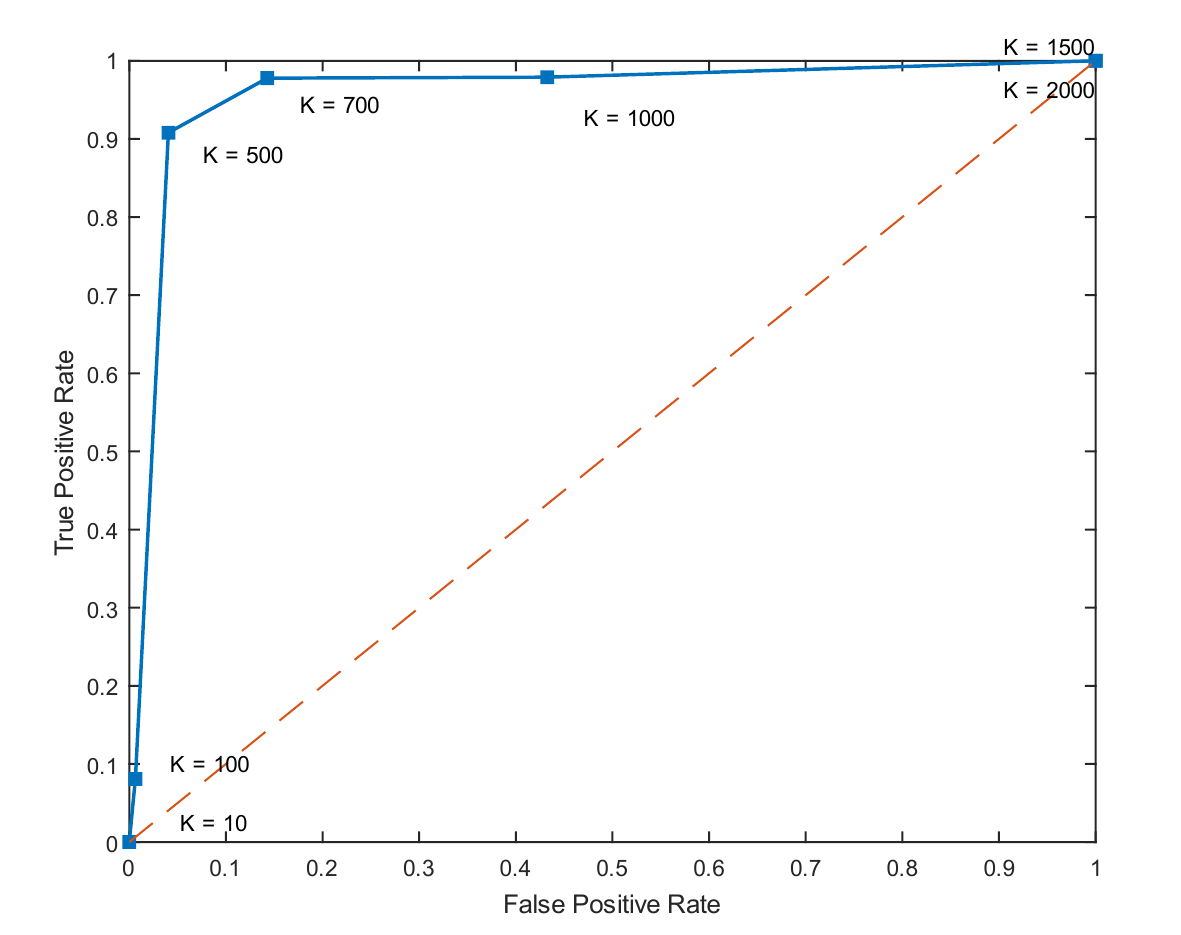}
    \caption{Receiver Operating Characteristic (ROC) curve for Algorithm 2 with the length of prediction horizon \(K\) as the third dimension.}\label{fig:results_validity_roc}
\end{figure}
\par
We can see from Figure~\ref{fig:results_validity_numsteps} that for a small prediction horizon length, Algorithm 2, returns mostly negative checks, with true and false negatives accounting for the vast majority of predictions for \(K < 500\). As \(K\) grows, the number of true and false positives increases, with the false positive rate increasing in a much slower manner. For \(K \geq 1500\), although the rate of true positives is high, Algorithm 2 returns a high number of false positives as well. This behaviour is the result of the design of Algorithm 2. First, for small \(K\), the algorithm will likely not be exploring a sufficient number of states where a stealthy attack would cause a violation of safety constraints. Thus, it is expected to observe a high rate of both false and true negatives, with true and false positive rates remaining very low. As \(K\) increases, the algorithm is allowed to explore more states, therefore increasing the number of true positives. The slow parallel increase in false positives shows that Algorithm 2 exhibits high accuracy for intermediate values of \(K\). However, at high values of \(K\), the accuracy of the predicted states is expected to decrease, which  explains the high false positive rates.
\par
These simulations  show that there exists a trade-off between how early we would like to raise warnings about potential safety violations due to a stealthy attack and the accuracy of Algorithm 2. These experiments can also serve as a method to tune the choice of \(K\). To showcase these ideas, we have plotted in Figure~\ref{fig:results_validity_roc} the Receiver Operating Curve (ROC) for Algorithm 2, with the length of prediction horizon \(K\) as the third dimension. We note that it exists a ``cut-off'' point at \(K=500\) time steps where we obtain acceptable values for the true/false positive rates (90.8\% for true positives, 4.05\% for false positives). This is equivalent to about \(15\) minutes ahead-of-time prediction, which is a reasonable choice in practice for \(K\).
\par
For  \(K= 500\), we tested the accuracy of Algorithm 2 under different numbers of sensors being attacked at the same time. We ran 500 simulations for each different number of sensors being attacked. In each simulation, we picked the attacked sensors at random, and we ran Algorithm 2 while considering the safety constraints associated with the sensor(s) under attack (Table~\ref{table:safety_constraints}). The results in Table~\ref{table:eval_pr_numSensorsAttacked} show high true positive and low false positive rates in each case. These experiments demonstrate the accuracy of Algorithm 2 with respect to all safety-critical sensors. 
Given the large number of random simulations we ran, we can conclude that Algorithm 2 can report potential safety violations due to a stealthy attack with respect to all the safety constraints imposed on the system.
\begin{table}[!t]
    \renewcommand{\arraystretch}{1.3}
    \caption{True positive/negative rates vs. the number of sensors attacked at the same time.}\label{table:eval_pr_numSensorsAttacked}
        \begin{center}
        \begin{tabular}{ c||c|c|c|c|c }
        \hline
        \textbf{Attacked Sensors} & \textit{1} & \textit{2} & \textit{3} & \textit{4} & \textit{5} \\
        \hline
        \textbf{True Positive Rate} & 0.908 & 0.905 & 0.91 & 0.905 & 0.92\\
        \hline
        \textbf{False Positive Rate} & 0.0405 & 0.041 & 0.04 & 0.0408 & 0.04\\
        \hline
        \end{tabular}
    \end{center}
\end{table}
\subsection{Comparison with Existing Monitoring Approaches}
In this section, we empirically showcase the usefulness of our approach compared to existing monitoring approaches described in Section~\ref{subsection:existing_monitoring_tech}. To this end, we implemented an online monitoring tool measuring the time-to-unsafe states metric based on the euclidean distance from the current state estimate to the set of unsafe states. We also measure the average rate of change of the estimated system state. We avoided a comparison based on the accuracy metrics depicted in the previous section. This is due to the fact that the traditional time-to-unsafe states metric relies on the selection of different thresholds to raise alarms of different criticality. With the lack of precise methods to select these thresholds, it is hard to perform a meaningful quantitative comparison between the metric proposed in this paper and the traditional one. Therefore, we used a set of attack scenarios  on safety critical sensors to empirically demonstrate the advantages of our approach. We particularly focus on attacks targeting sensors with a slowly growing bias.
\par
\begin{figure*}[!t]
    \centering
    \includegraphics[scale = 0.7, angle = 0]{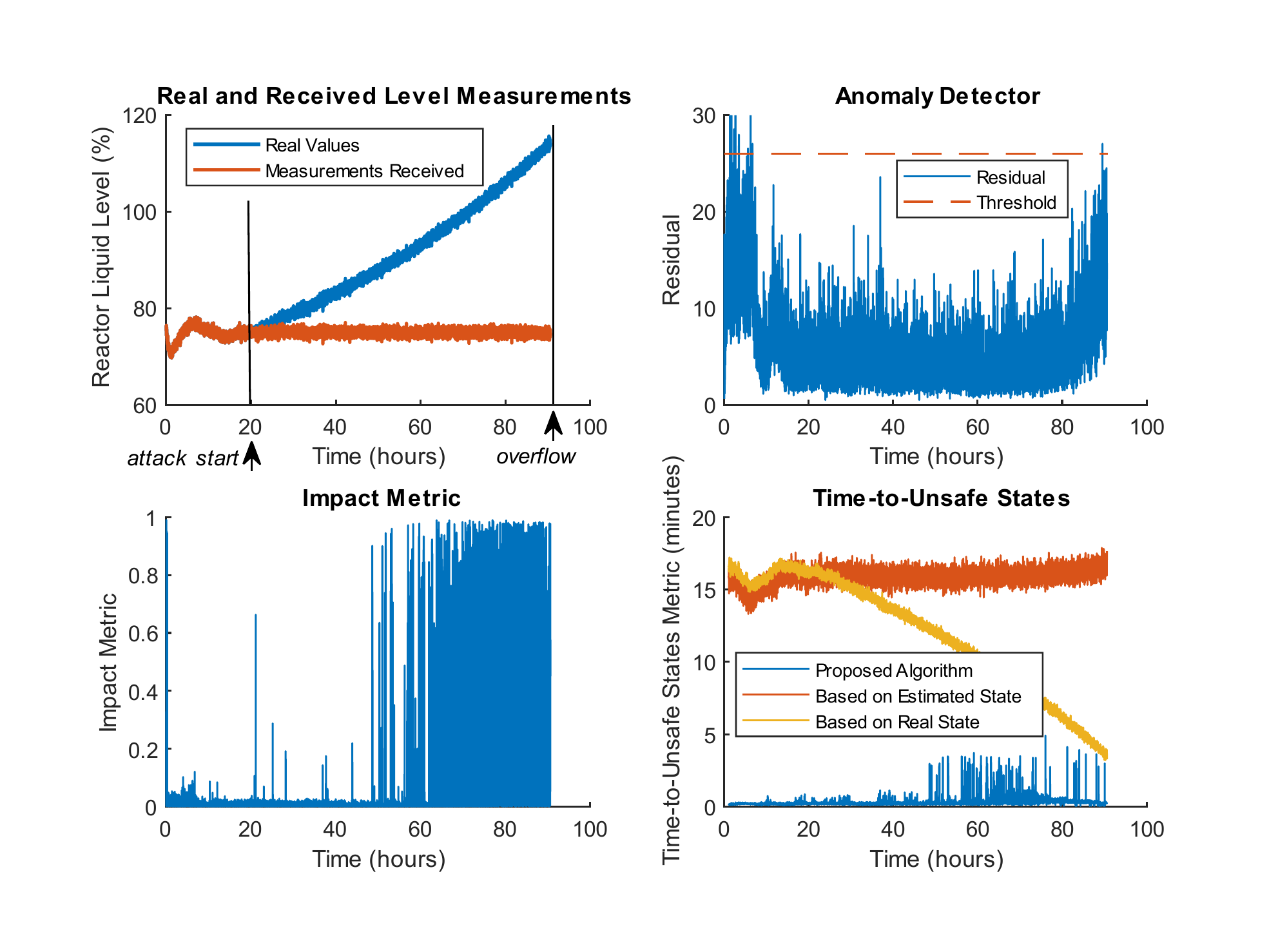}
    \caption{Results from Simulation Scenario 1.}\label{fig:results_scenario1}
\end{figure*}
\begin{figure*}[!t]
    \centering
    \includegraphics[scale = 0.7, angle = 0]{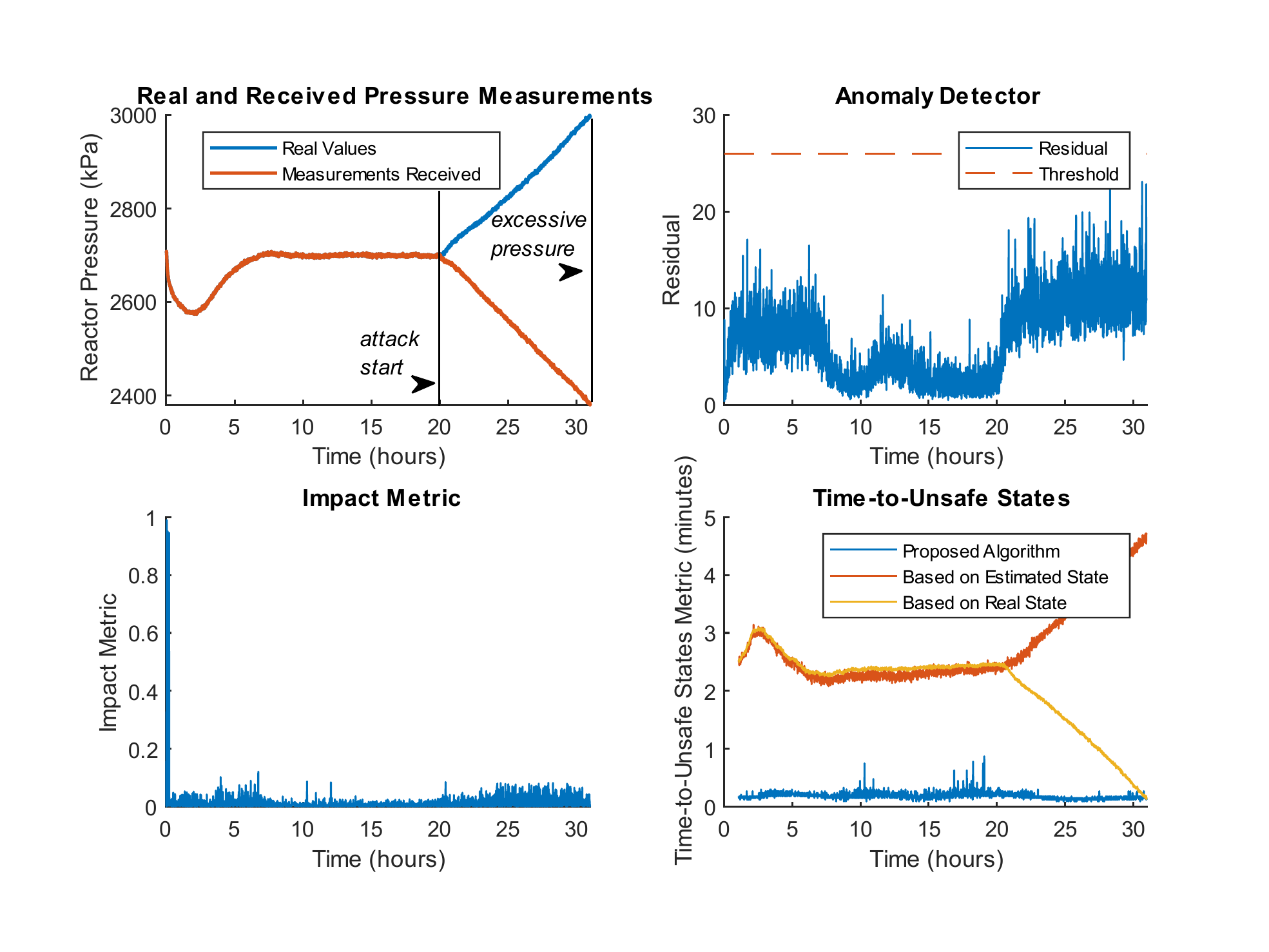}
    \caption{Results from simulation Scenario 2.}\label{fig:results_scenario2}
\end{figure*}
\begin{figure*}[!t]
    \centering
    \includegraphics[width = \textwidth, keepaspectratio, angle = 0]{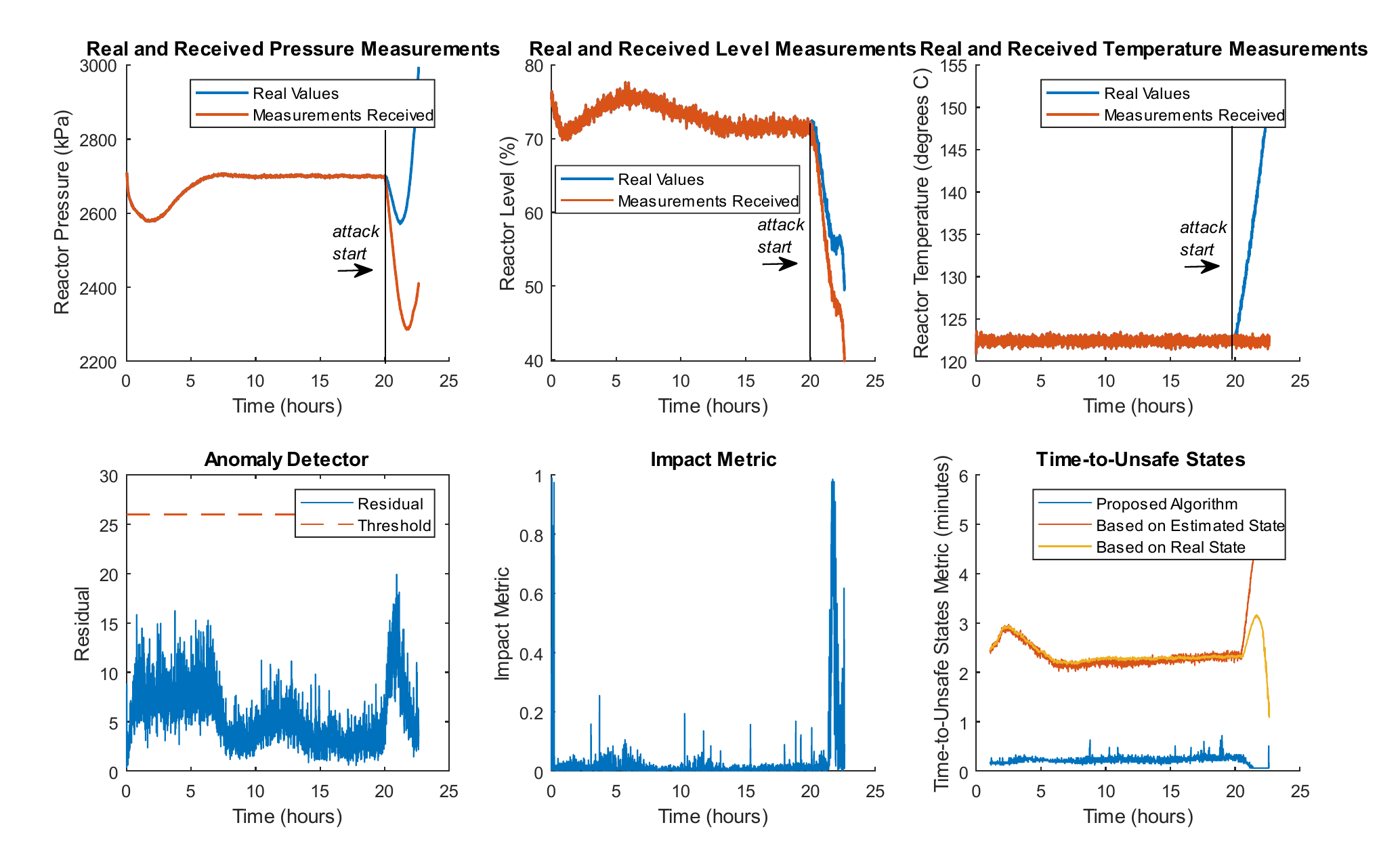}
    \caption{Results from simulation Scenario 3.}\label{fig:results_scenario3}
\end{figure*}

We chose three attack scenarios. Scenarios 1 and 2 depict individual attacks targeting the level and pressure sensors, respectively, of the main reactor in the TEP. Scenario 3 depicts an attack performed simultaneously on the level, pressure, and temperature sensors of the reactor. We chose Scenarios 1 and 2 to illustrate the typical kind of attacks targeting safety-critical sensors, individually. We can obtain similar results for individual attacks on other safety-critical sensors. Scenario 3 illustrates a more dangerous coordinated attack where all main reactor sensors in the TEP are manipulated at the same time. Again, similar results can be obtained for other combination of sensors for safety-critical variables listed in Table~\ref{table:safety_constraints}. Figures~\ref{fig:results_scenario1},~\ref{fig:results_scenario2} and~\ref{fig:results_scenario3} present the results obtained for each scenarios.
\par
\subsubsection{Scenario 1} In this scenario, we simulate an attack on the reactor's level sensor, where a growing bias on level measurements is introduced to trick the controller into overflowing the reactor. This simulation is shown in Figure~\ref{fig:results_scenario1} where we can see that the anomaly detector raises an alarm almost at the moment the overflow takes place. Our impact metric increases significantly over the period preceding the physical damage to the reactor, and our time-to-unsafe states metric shows that the plant is dangerously close to damage. Conversely, the traditional time-to-unsafe states computed based on the estimated state alone shows a slight increase, indicating that the plant appears to be moving away from the unsafe operating region.
\subsubsection{Scenario 2} In this scenario, we perform an attack on the reactor's pressure sensor. Namely, the attack is a slowly growing bias on pressure measurements seeking to trick the pressure controller into increasing the pressure inside the reactor while it appears lower than its set-point. We can see from Figure~\ref{fig:results_scenario2} that the anomaly detector fails to raise any alarm before excessive pressure builds up in the reactor. However, our online monitoring algorithm reports that the plant's operation may be unsafe in the presence of an attack throughout this ramp-down operation, which is shown by a non-zero impact metric. While the impact metric shows an increase over the few hours between the start of the attack and the damage taking place. Instead, the traditional time-to-unsafe states metric shows the plant moving away from unsafe states.
\subsubsection{Section 3} In this scenario, we simulate simultaneous attacks on the main reactor's pressure, temperature, and level sensors. All three attacks are slowly growing biases. Figure~\ref{fig:results_scenario3} shows that damage occurs faster in this scenario than in the previous two, with the anomaly detector again failing to raise any alarms. 
Our impact metric however shows again that the plant's operation may be unsafe under the attack. Instead, the traditional time-to-unsafe states metric depicts the plant moving away from the unsafe operating region.
\par
These scenarios demonstrate the usefulness of our approach in the presence of stealthy attacks when compared to simple distance-to-unsafe metrics. Relying on the traditional distance-to-unsafe metric may relay an inaccurate idea of the current security or safety conditions. This was especially highlighted in Scenarios 2 and 3. While the plant appears to drift away from the unsafe operating region, our monitoring approach can still warn operators that an attacker is able to damage the system without being detected.
\par
\begin{figure*}[!t]
    \centering
    \includegraphics[width = \textwidth, keepaspectratio, angle = 0]{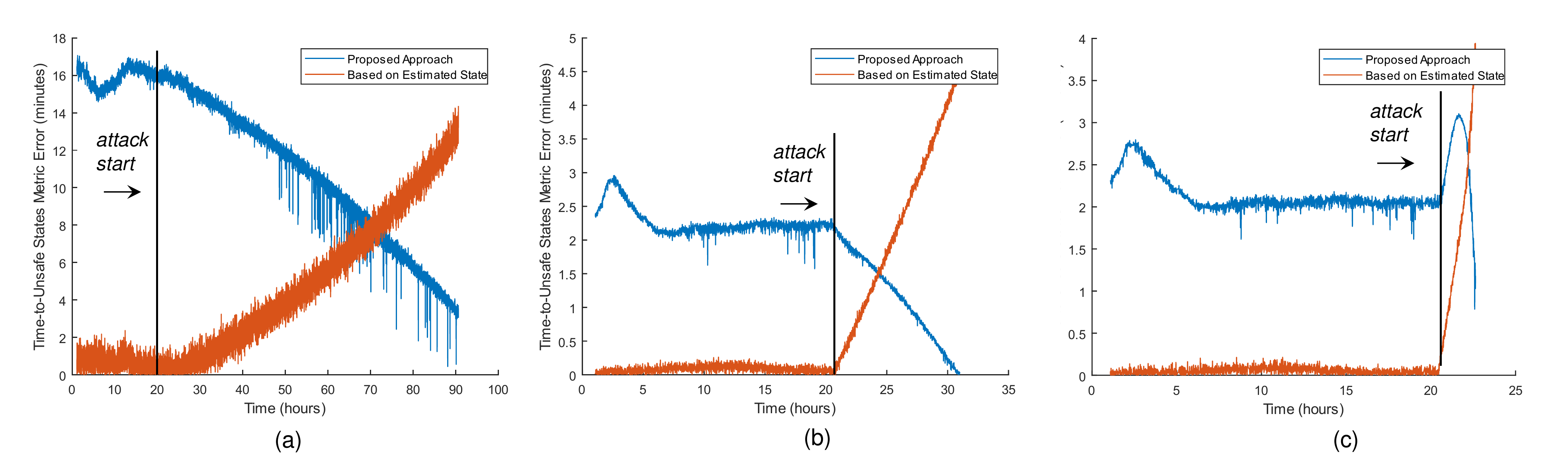}
    \caption{Difference (error) between time-to-unsafe metric computed based on state estimate and based on the proposed algorithm vs. the time-to-unsafe states based on the real state of the system; (a) - Scenario 1, (b) - Scenario 2, (c) - Scenario 3.}\label{fig:results_scenarios_compare}
\end{figure*}
\par
Figure~\ref{fig:results_scenarios_compare} shows a comparison between the time-to-unsafe metric computed using our algorithm and the same metric computed based on the raw estimated state. Namely, we plot the difference (error) between the metric in each case and the time-to-unsafe states computed based on the real state of the system. In each scenario, we observe that  the metric based on the raw estimated state is relatively accurate before the attack starts (the error is close to zero). However, the error starts to grow as the stealthy attack progresses and the real state diverges from its estimate. Conversely, as the stealthy attack progresses, this error decreases for the time-to-unsafe states metric computed according to our algorithm and reaches almost zero towards the end of the attack.
\par
This demonstrates the usefulness of our algorithm in the worst  case, where the actual state of the system significantly diverges from the real estimate under a stealthy attack. While this may be overly conservative when the system is not under attack, the safety criticality of the systems we consider justifies the need to employ a monitoring that can generate early warnings when a dangerous stealthy attack is taking place. Hence, the task of pre-computing reachable sets under attacks and using them for online monitoring is well justified.
\par
Our  monitoring approach can be used as part of an early warning system to improve situational awareness and potentially preserve important data relevant to an investigation should a stealthy attack indeed cause damage. A full treatment of this aspect of our approach is however deferred for future work.

\subsection{Performance and Scalability}
\begin{figure}[!t]
    \centering
    \includegraphics[width=\columnwidth,keepaspectratio]{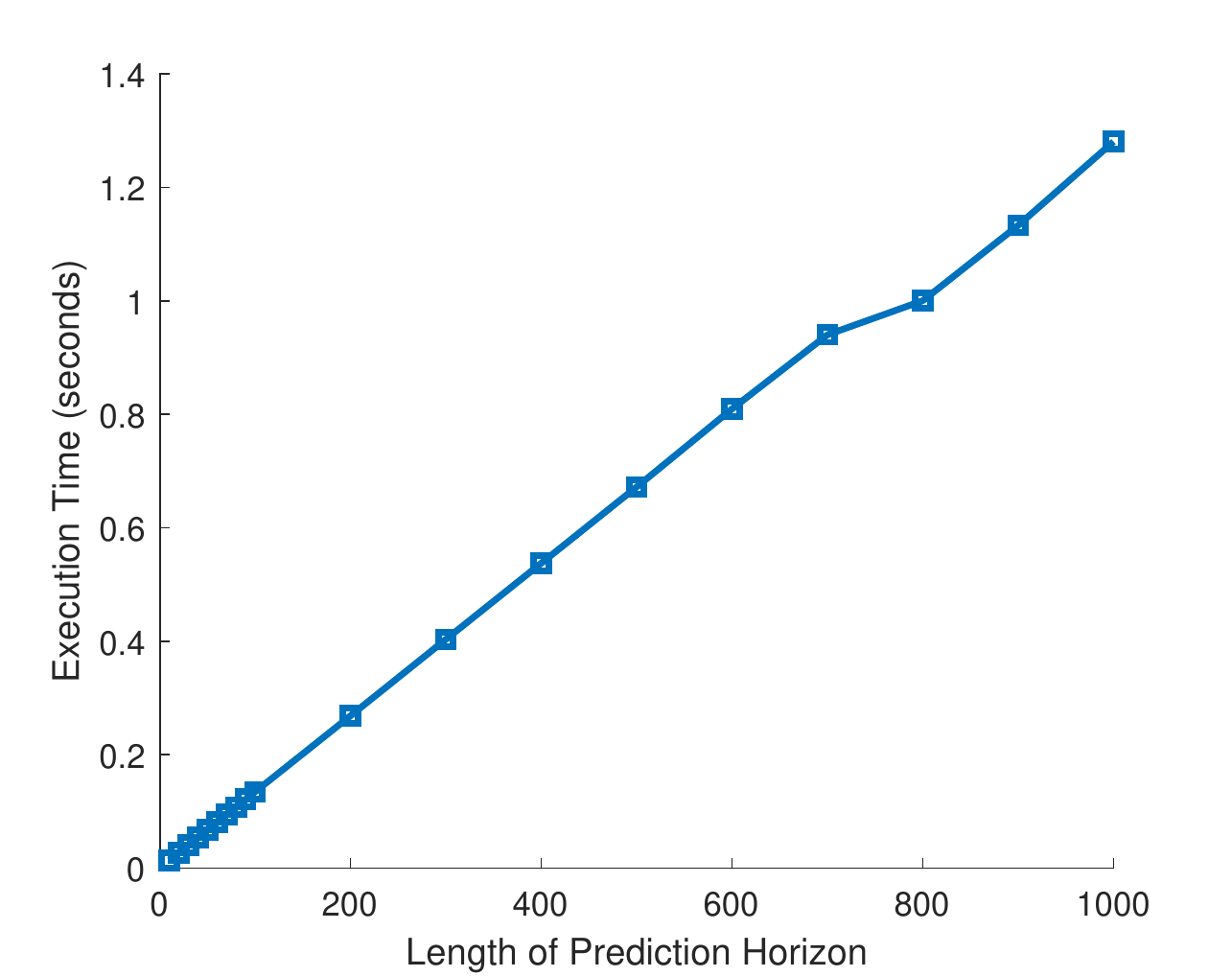}
    \caption{Average execution time of the proposed scheme vs.\ the number of steps for prediction.}\label{fig:results_perf}
\end{figure}
We ran \(100\)-hour simulations of the TEP to test the performance of our approach. For performance testing purposes, we modified the algorithm so that all the states along the prediction horizon are visited. This represents the worst-case  scenario. We tested the ability of Algorithm 2 to scale with respect to (i) the length of the prediction horizon (number of steps \(K\)), and (ii) the number of safety constraints. We performed this testing  on a machine with an Intel i7-9750H CPU clocked at 2.6 GHz and with 16 GB of RAM.
\par
Figure~\ref{fig:results_perf} shows the average time needed to perform state prediction and safety checking (Algorithm~\ref{alg:online}) for each number of steps, with a fixed number of safety constraints~\footnote{For a \(100\)-hour simulation, this average is taken over approximately \(2\times10^{5}\) safety checks for a sampling period of \(1.8\) sec.}. Results show that the performance of Algorithm 2 scales linearly with the number of steps required for state prediction. Furthermore, at \(K = 1000\) steps, equivalent to approximately \(30\) min ahead-of-time prediction, the execution time of the safety checking algorithm is smaller than the sampling period (\(1.8\) sec).
\par
\begin{figure}[!t]
    \centering
    \includegraphics[width=\columnwidth,keepaspectratio]{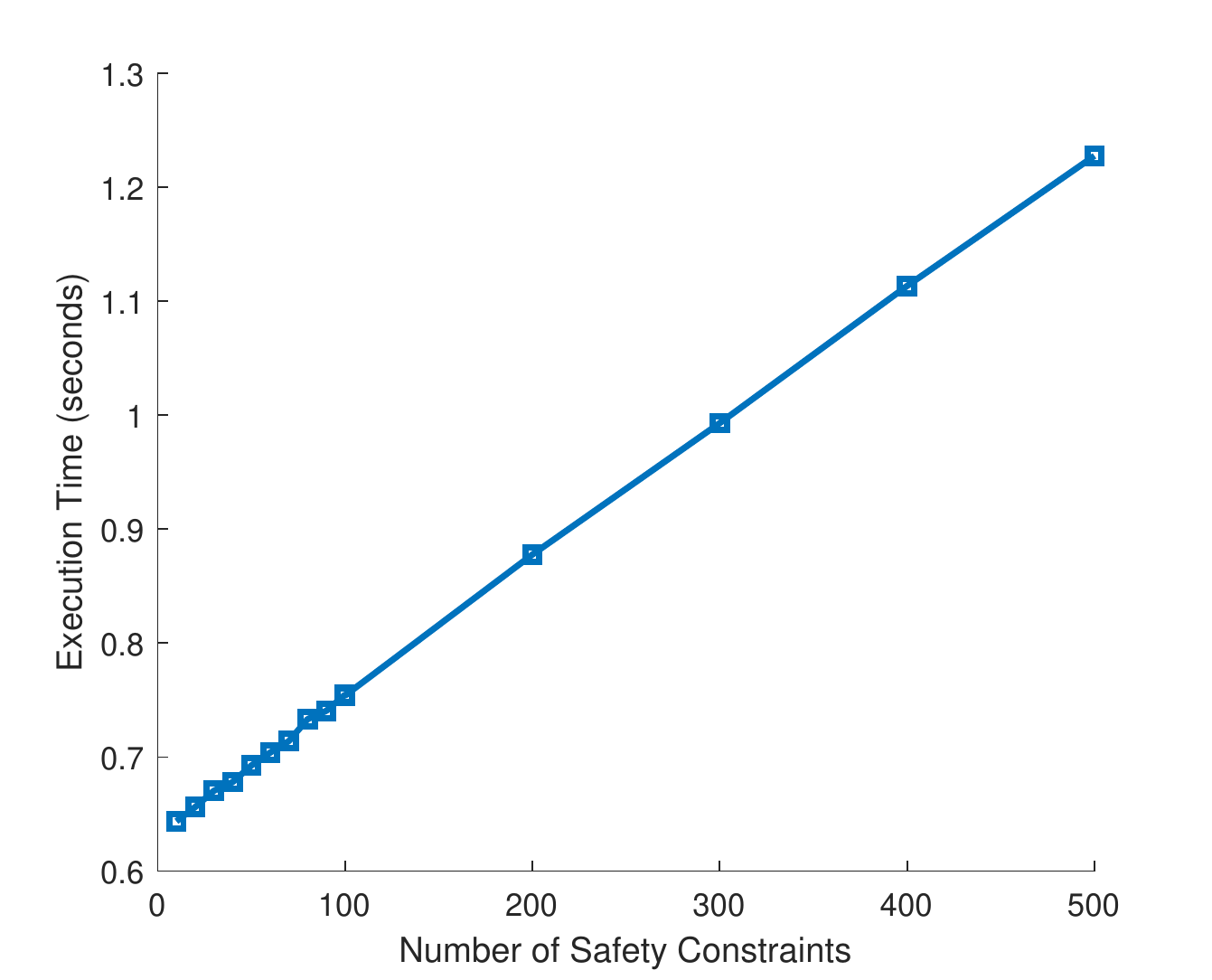}
    \caption{Average execution time of the proposed scheme vs.\ the number of safety constraints.}\label{fig:results_perfsafetyconst}
\end{figure}
In addition, we tested the ability of our algorithm to scale with the number of safety constraints. To this end, we fixed \(K\) at \(500\) steps (i.e. \(\approx 15\) min) and we ran \(100\)-hour simulations for each number of safety constraints. For the purposes of testing, we generated random half-spaces representing safety constraints. Results are shown in Figure~\ref{fig:results_perfsafetyconst}. The execution time scales linearly with the number of safety constraints, and with 500 safety constraints is still less than the sampling time of the system. Hence, the proposed algorithm exhibits excellent real-time performance in the presence of more complex safety constraints.
\par
It is worth noting  that the performance of the proposed algorithm can be  improved significantly if implemented with a compiled language such as C++ instead of MATLAB. This is the case in most control systems applications.

\section{Conclusion}\label{section:conclusion}

In this paper we have presented a predictive online safety monitoring
approach for LTI systems under potential stealthy sensor attacks. Our approach precomputes offline symbolic reachable
sets in terms of the system's state estimate, by considering the
evolution of the estimation error under a potential stealthy
attack. Given the current state of the system and
controllers, we predict in real time the control flow of the system for a certain
number of steps in the future. The precomputed sets are then
instantiated at the predicted estimates. We use ellipsoidal calculus techniques to perform emptiness checks of the intersection of the precomputed set with a set of unsafe states. We applied the approach to the large-scale Tennessee-Eastman process (TEP) where we validated our approach and we showed that it can perform safety checks in a timely manner. Furthermore, we demonstrated the improvement over existing online monitoring techniques and we showed that the computation of reachable sets under stealthy attacks is well justified in safety-critical applications.
In the future, we will study in more detail the uncertainty propagation caused by the prediction of future states and its effect on the validity of the safety checking.

\printbibliography{}
\end{document}